\documentclass[prl, twocolumn, showpacs, floatfix, letterpaper, groupedaddress]{revtex4}
\usepackage[dvips]{graphicx}
\begin{document}

\title{The depletion force in a bi-disperse granular layer}
\author{Paul Melby}
\altaffiliation{Present Address: Qbit, LLC, 6905 Rockledge Dr, 3rd Floor, Bethesda, MD 20817, USA}
\author{Alexis Prevost}
\altaffiliation{Present Address: Laboratoire de Physique Statistique de l'Ecole Normale Sup\'erieure, CNRS-UMR 8550, 24 rue Lhomond 75231 Paris cedex 05, France\\}
\author{David A. Egolf}
\author{Jeffrey S. Urbach}
\altaffiliation{Corresponding Author}
\email[]{urbach@physics.georgetown.edu}
\affiliation{Department of Physics, Georgetown University, 37th \& O Streets, Washington DC
20057, USA}

\date{\today}


\begin{abstract}

We demonstrate the effect of the depletion force in experiments and simulations of vertically vibrated mixtures of large and small steel spheres. The system exhibits size segregation and a large increase in the pair correlation function of the large spheres for short distances that can be accurately described using a combination of the depletion potential derived for equilibrium colloidal systems and a Boltzmann factor. The Boltzmann factor defines an effective temperature for the system, which we compare to other measures of the temperature.

\end{abstract}

\maketitle

Vibrated granular materials show many complex behaviors, including pattern formation, segregation, non-equipartition of energy, and non-Gaussian velocity distributions \cite{jaeger, kadanoff, melby04}.  Many of these are purely nonequilibrium effects, but some features of vibrated granular materials are strikingly similar to equilibrium systems such as colloidal suspensions \cite{melby04, prevost04, olafsen05}.  The extent to which the machinery of equilibrium statistical mechanics can be used as a starting point to develop a theory of these non-equilibrium steady states remains an open question.

Segregation mechanisms, meanwhile, are important in many industrial applications.  Granular mixtures are known to segregate by size, mass, shape, and frictional coefficients and this segregation can be caused by many mechanisms, including vibration, convection, and tumbling \cite{kudrolli, ottino}.  In this Letter, we investigate a relatively unexplored mechanism for segregation in granular materials: the depletion force.  Well studied in colloidal systems \cite{asakura, russel, frydel05}, the depletion force is an entropic effect in which the introduction of small particles into a colloidal dispersion will cause the larger colloidal particles to segregate.  Previous suggestions of the role of the depletion force in granular materials have focused on horizontally shaken \cite{mullin, reismullin, reis04} or horizontally swirled \cite{aumaitre} mixtures.  The segregation effects for the horizontally vibrated mixtures, however, has been shown to be caused by a differential driving mechanism\cite{ehrhardt04}, which may also hold for the horizontally swirled system.  In this paper we show that the depletion force can cause size segregation in a vertically vibrated, bi-disperse granular layer.  The pair correlation function for low concentrations of the large particles can be accurately described using the depletion potential derived for equilibrium systems and a Boltzmann factor.  This Boltzmann factor defines an effective temperature for the system, which we compare to other measures of temperature in the system.

\begin{figure}[!ht]
\includegraphics[width=7cm]{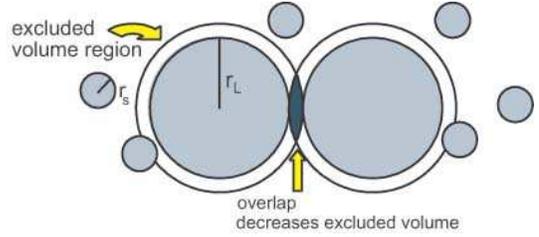}
\caption{Schematic drawing of the depletion interaction.  The white shell around the large particles represents the volume from which the small particles are excluded.  The dark blue region shows the overlap in this excluded volume when the large particles are close together.} 
\label{fig:depletion}
\end{figure}

The depletion force was originally proposed by Asakura and Oosawa in 1954 \cite{asakura}.  There are two approaches for deriving the potential: the first is thermodynamic and the second is mechanical.  In the thermodynamic view, illustrated in Fig. \ref{fig:depletion}, the depletion force is derived through entropy maximization.  There is a shell of width $r_{s}$ (the radius of the smaller particles) around the large spheres into which the centers of the small spheres cannot penetrate.  By arranging two large spheres together, the excluded volume regions of the two large spheres overlap, and the volume accessible to the small spheres increases, increasing the total entropy of the system.  This produces an effective attraction between the large spheres.  The mechanical view of the depletion force is that the small spheres exert a pressure on the large spheres, but that pressure is unbalanced when two large spheres are close enough together that the small spheres are excluded from the space between the spheres.  The net force is found by integrating the pressure over the surface of the large spheres.  In equilibrium, both views are equivalent and give the following expression for the depletion potential for two large spheres separated by a distance $r$, valid for moderately low densities of small spheres:
\begin{equation}
U(r) = \left\{ \begin{array}{ll} \infty & r <  2r_{L} \\
  P V_{ex}& 2r_{L} < r < 2r_{L} + 2r_{s}\\
   0 & 2r_{L} + 2r_{s} < r
\end{array} \right. 
\end{equation}
where $r_{L}$ is the radius of the large spheres,  $P$ is the pressure exerted on the large spheres by the small spheres and $V_{ex}$ is the overlap in excluded volume.  $V_{ex}$ can be expressed as:
\begin{equation}
 V_{ex}(r) = \frac{-4 \pi (r_{s} +r_{L})^3}{3} \left[ 1 - \frac{3 r}{4(r_{L}+r_{s})} + \frac{r^3}{16(r_{s} + r_{L})^3} \right] .
 \label{eq:depletion}
 \end{equation}

\begin{figure}[!ht]
\includegraphics[width=4cm]{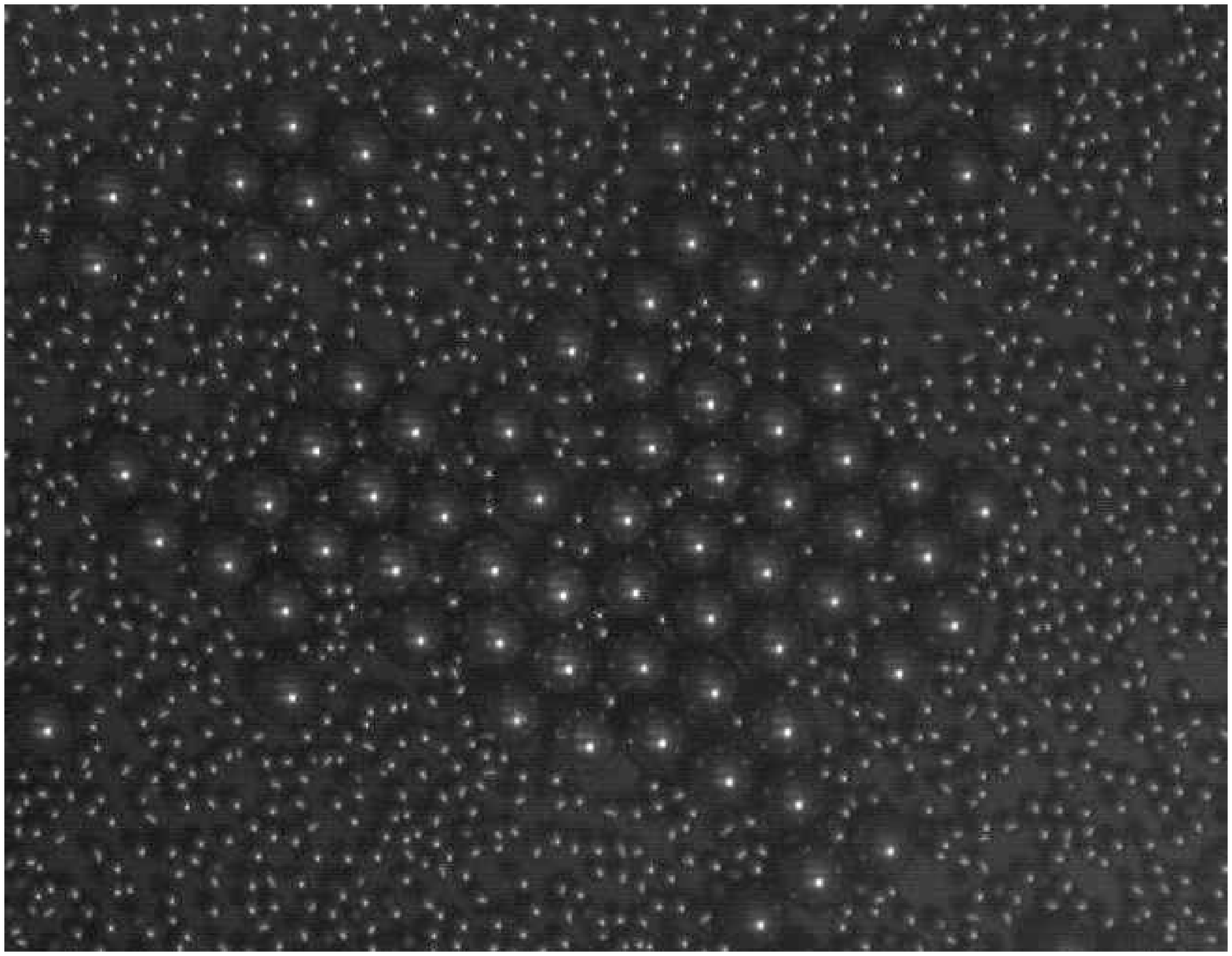} 
\includegraphics[width=4.15cm]{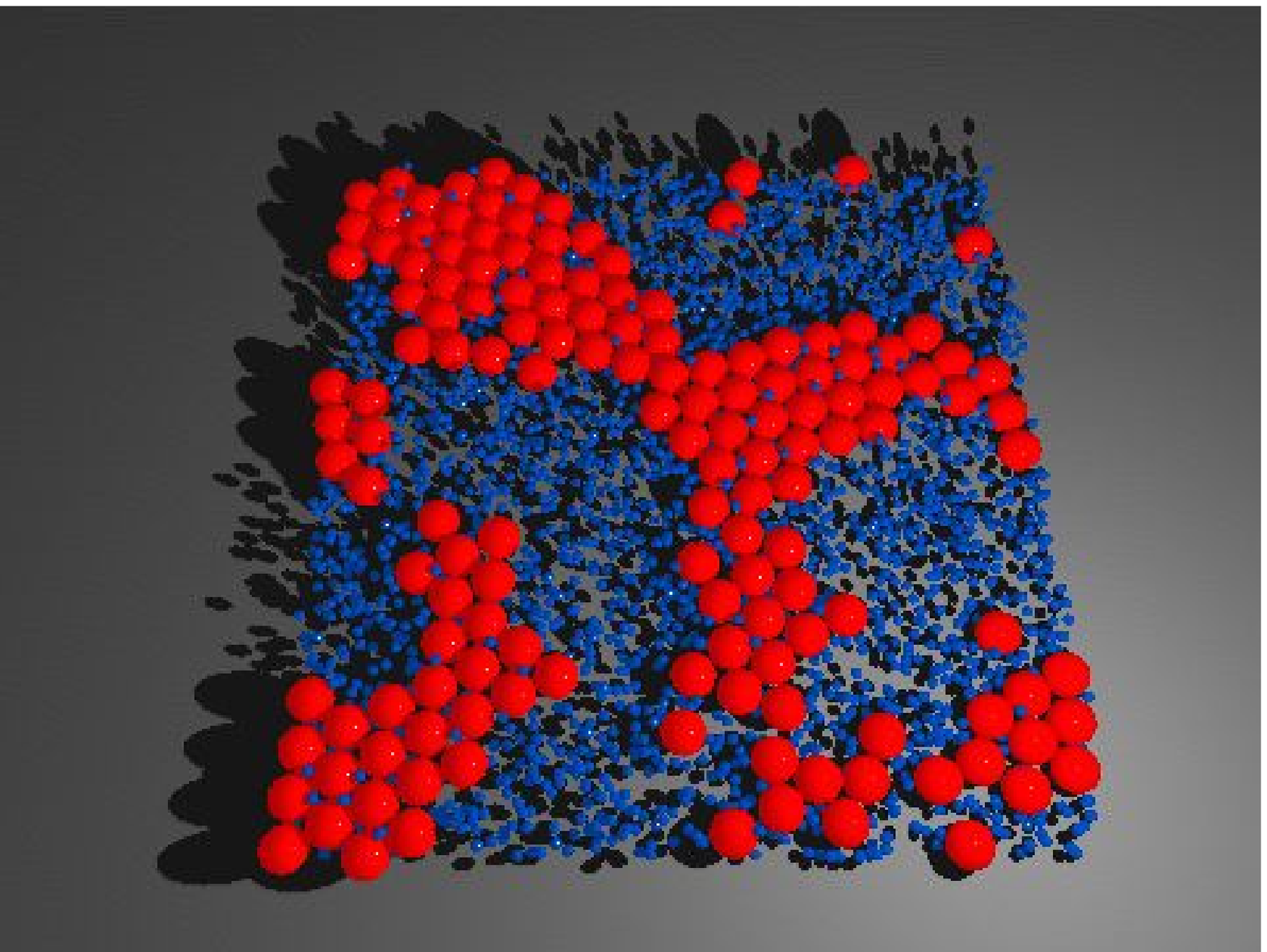}
\caption{Segregation due to the depletion force. Left panel: Experiment ($\rho _{s} = 0.036$, $\rho _{L} = 0.066$). Right Panel:  Simulations ($\rho _{s} = 0.080$, $\rho _{L} =  0.132$).  In both instances the large and small spheres were initially randomly distributed throughout the available space.  $\nu = 30$ Hz, $\Gamma = 3.5$,   $r_{L}=4$ mm, $r_{L}/r_{s}$=3.33, $h = 2.3 r_{L}$. } \label{fig:segregation}
\end{figure}

Our experimental setup consists of steel spheres of two sizes placed between two horizontal plates which are vibrated sinusoidally in the vertical direction.  Images captured by a high resolution digital camera  are analyzed to find the positions of the large spheres.   For a range of ball diameters, as well as a range of both large and small ball volume fractions, we observe segregation of the large spheres.  Fig. \ref{fig:segregation}a shows the results of one such experiment, where large spheres initially distributed randomly on the plate have formed a compact cluster.  The parameter values, including the gap between the bottom and top plates, $h$, the vibration frequency, $\nu$, the dimensionless acceleration $\Gamma = A (2 \pi \nu)^2/g $, where $A$ is the vibration amplitude, and  the volume fraction of large and small spheres, $\rho _{L,s}=N_{L,s}V_{L,s}/V$, where $V_{L,s}$ is the volume of the large or small spheres and $V$ is the volume of the system, are given in the figure caption. The segregation behavior is relatively insensitive to the frequency and amplitude of the acceleration, but is sensitive to the gap spacing.  For spacings very close to the large ball diameter, friction with the top and bottom plates severely limits motion of the large spheres.  For spacings significantly larger than the large sphere diameter, the kinetic energy of the large spheres is much larger than that of the small spheres, and the forces from the small spheres do not appreciably affect the trajectories of the large spheres.  This strong deviation from equipartition is a consequence of the strong interaction with the vibrating plate experienced by all of the particles in the thin granular layers.  Except as noted, all of the results presented below were taken with the parameters used for Fig.~\ref{fig:segregation}.

Because the system is highly sensitive to variations in the flatness of the plates, we also performed molecular dynamics simulations of the system to ensure that the segregation was not caused by such experimental issues, using a model that has been shown to reproduce the dynamics of similar granular experiments \cite{nie, prevost, prevost04, melby04}.  The coefficient of restitution used in the simulation, which is set by a combination of the elastic restoring force and the dissipation, was 0.9.  The other parameters were chosen to match the experiment. The simulations produced phase segregation similar to that observed in the experiment, one example of which is shown in Fig. \ref{fig:segregation}.  

\begin{figure}[!ht]
\includegraphics[width=6.75cm]{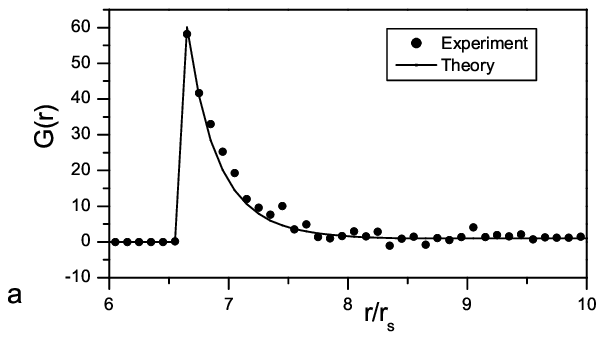}
\includegraphics[width=6.75cm]{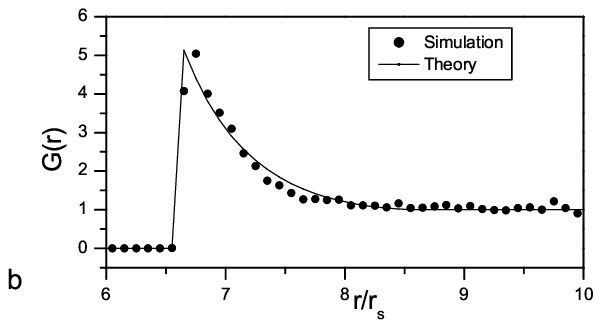}
\includegraphics[width=6.0cm]{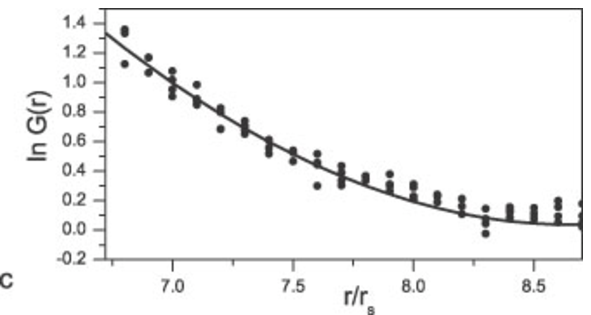}
\caption{Pair correlation function of large particles and the fit using the a Boltzmann factor and the equilibrium depletion potential (Eq. \ref{eq:depletion})  a) Experiment ($\rho _{s} = 0.071$, $\rho _{L} = 0.007$) b) Simulations ($\rho _{s} = 0.08$, $\rho _{L} = 0.03$) c) Logarithm of G(r) for several simulation runs ($\rho _{s} = 0.06$, $\rho _{L} = 0.026$).  Solid lines are fits to the equilibrium theory (see text). 
} \label{fig:paircorrelation}
\end{figure}

In equilibrium hard sphere systems, the depletion force can be measured directly by looking at the pair correlation function.  If the density of large spheres is small enough that only two-particle interactions are present, then the pair correlation function, $G(r)$, can be simply described through the use of a Boltzmann factor: $G(r) = \exp(-U(r)/{T})$, in units where $k_B=1$.  In the non-equilibrium granular system, there is no reason to expect that this equilibrium approach should accurately describe the pair correlation function.  Fig. \ref{fig:paircorrelation}a shows an example of  the pair correlation function measured in the experiment, and a result from the simulation is shown in  Fig. \ref{fig:paircorrelation}b.  The solid curve is a fit to $G(r) = \exp(-V_{ex}P/{T})$, with  $V_{ex}$ given by Eq. \ref{eq:depletion} and $P/{T}$ a fitting parameter that provides only an overall scale factor.  The equilibrium-like approach  accurately reproduces the width of the peak in $G(r)$ and the overall shape of the curve.  We performed simulations with the small sphere radius reduced by 60\% and found similar agreement.
The fact that the range of the interaction between the large spheres is set by the small sphere radius is unequivocal evidence that the attraction is due to the depletion force.  

\begin{figure}[!ht]
\includegraphics[width=6.75cm]{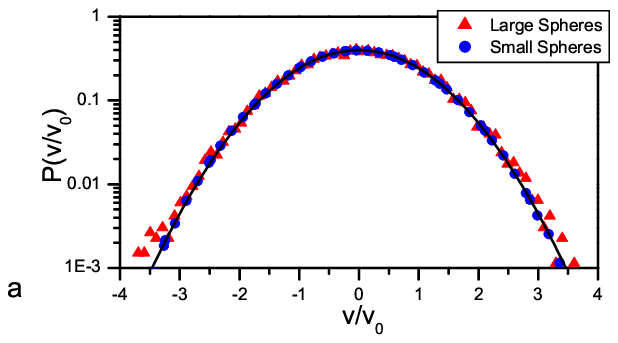}
\includegraphics[width=6.75cm]{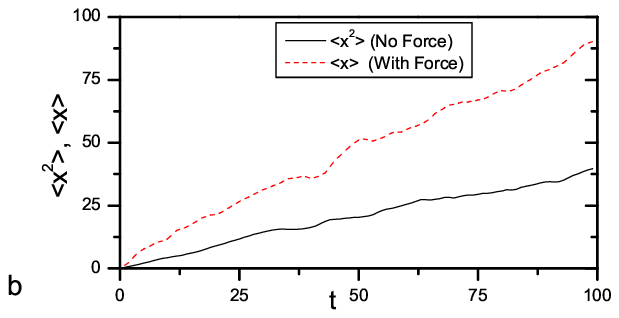}
\includegraphics[width=6.0cm]{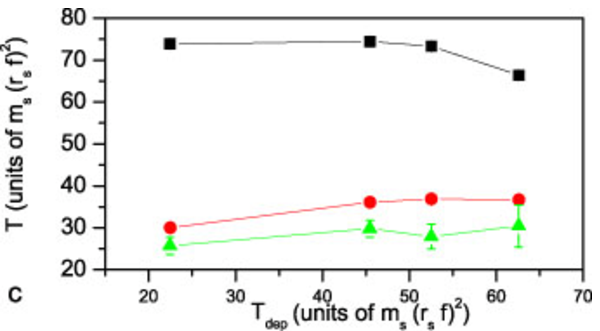}
\caption{Temperatures measured in the simulations of the granular system. a) Velocity distributions of small and large particles in the simulations with the same parameters as in Fig. \ref{fig:paircorrelation}c ($v_0=\sqrt{<v^2>})$. The solid line is a Gaussian distribution.  (b) Average displacement $<x>$ as a function of time for a large particle subjected to a constant external force, and mean squared displacement  $<x^2>$ vs. time for an unforced particle  ($\rho _{s} = 0.096$, $\rho _{L} = 0.0026$). (c) The granular temperatures of the small (black squares) and large (red circles) spheres and the fluctuation-dissipation temperature (green triangles) at different densities of the small spheres plotted vs. the  
effective temperature measured from for the depletion potential  (From lowest to highest $T_{dep}$: $\rho_s$=0.036, 0.060, 0.072, 0.097.)
}
\label{fig:alltemps}
\end{figure}

In the simulations, the pressure in the system can be calculated by use of the virial theorem \cite{prevost04, rapaport}, allowing an explicit calculation of an `effective temperature', $T_{dep}$, from the value of $P/{T}$ obtained from the fit to $G(r)$.   In order to determine if this temperature is related to other fluctuations in the bi-disperse system, we have measured three other effective temperatures in the computer simulations, the granular temperature of the small spheres, the granular temperature of the large spheres, and an effective temperature defined through a fluctuation-dissipation relation.  The granular temperatures are given by $m <v_i>^2$, where $v_i$ is one component of the particle velocity and $m$ is its mass.  The distributions of the horizontal velocity components for the large and small particles are shown in Fig. \ref{fig:alltemps}a.  Both are approximately Gaussian for small values of the velocity, and the small particles are well described by a gaussian distribution over all of the accessible range.  The smaller particles are less confined and hence have a more three-dimensional motion, which has been shown to lead to more nearly Gaussian velocity distributions \cite{Olafsen99}.

The granular temperatures are compared to $T_{dep}$ for several concentrations of small spheres in Fig. \ref{fig:alltemps}c.  The results show that, unlike either granular temperature, $T_{dep}$ is a strong function of small sphere density.  One possible explanation for this behavior is that the pressure in the vicinity of the large balls differs significantly from the average pressure due to non-equilibrium effects.
We also investigated whether $T_{dep}$ is related to an effective temperature determined by a fluctuation-dissipation relation.    While the granular system is not in the linear non-equilibrium regime, some studies have found that fluctuation-dissipation measurements of the temperature can agree with the granular temperature \cite{puglisi, barrat}.  In particular, in Ref. \cite{barrat}, it was found that the granular temperature of each species of particles in a binary granular gas matches the temperature calculated by applying the Einstein relation for that species.  The Einstein relation connects the coefficient of viscous drag, $k_{\rm{drag}}$, to the diffusion constant, $D$,  through the temperature: $T=k_{\rm{drag}} D$.  In the simulations, we can apply a constant external force to a large sphere and measure $k_{\rm{drag}}$ using the relation   $F=k_{\rm{drag}}v_{\rm{term}}$, where $v_{\rm{term}}$ is the terminal velocity.  An example of the linear displacement vs. time in the presence of a constant external force is shown in Fig.  \ref{fig:alltemps}b.  We can also measure the diffusion constant in the case of no external forces using the relation   $<x^2> = 2Dt$. An example of this is also shown in Fig.  \ref{fig:alltemps}b. Combining these measurements gives us an independent measurement of  temperature in the system:  $T_{FD} = D F_{ext}/v_{\rm{term}}$.
As can be seen in Fig.  \ref{fig:alltemps}c, the fluctuation-dissipation measurement of the temperature does not match the depletion temperature.  Also, unlike the results described in Ref. \cite{barrat}, there is no agreement between the fluctuation-dissipation temperature and either granular temperature, although it is relatively close to the value found for the large spheres.  This difference may be due to the forcing or dissipation from the top and bottom boundaries, which were not present in the model of Ref. \cite{barrat}.   Similarly, in quasi-2D colloidal systems, the presence of the confining plates results in some small additional contributions to the depletion interaction \cite{frydel05} that have not been considered in the above calculations.

\begin{figure}[!ht]
\includegraphics[width=7cm]{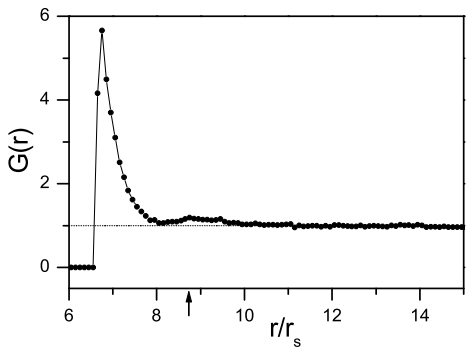}
\caption{Pair correlation function for large spheres in the case of a high concentration of small spheres ($\rho _{s} = 0.12$, $\rho _{L} = 0.026$).  The smaller second peak in G(r) occurs at $r=2r_{L} + 2r_{s}$ (indicated by the arrow), where exactly one small sphere fits in between the large spheres. }
\label{fig:extrapeak}
\end{figure}

As the density of small balls increases, the simple form of the depletion potential shown in Eq. \ref{eq:depletion} is no longer accurate in equilibrium systems \cite{deplete-ref}.  From an entropic point of view, the most favorable separations are those that allow integer numbers of small spheres to fit between large spheres.  This leads to additional peaks in the pair correlation function in the equilibrium case at integer multiples of the small sphere diameter.  This many-body effect, which is not easy to explain from the mechanical arguments, shows up as a small peak in $G(r)$ for $r=2r_L + 2r_{s}$.  This peak is visible in the pair correlation function determined from a simulation at a high density of small spheres, as shown in Fig. \ref{fig:extrapeak}.   This demonstrates that subtle entropy-driven dynamics are observable in this  far from equilibrium granular system.  

In summary, we have found direct evidence for the depletion force in a vibrated granular system.  We observe segregation of large particles  induced by the presence of small particles, and an increase in the pair correlation function of the large particles at low concentrations which is consistent with the equilibrium depletion potential.  We also observe, for high concentrations of small particles, a second peak in the large particle pair correlation function that is characteristic of the depletion potential.  Using an approach from equilibrium statistical mechanics, we measure an effective temperature for the system using the depletion potential and a Boltzmann factor.  This effective temperature does not agree with the granular temperatures of either sized particle or with a measurement of the temperature with a simple fluctuation-dissipation measurement.  These observations suggest that the driving mechanisms of the depletion interaction can operate in far from equilibrium granular systems, but that the equilibrium tools for quantifying the effects of the depletion force need modification to account for nonequilibrium effects such as energy injection and dissipation.

\begin{acknowledgements}
We have benefitted from helpful discussions with Paul Umbanhowar and Pedro Reis.  This work was supported by The National Science Foundation with grants DMR-9875529 and DMR-0094178 and by NASA under award number NNC04GA63G.  D.A.E. is also
supported as an Alfred P. Sloan Research Fellow. 
\end{acknowledgements}


\begin{thebibliography}{99}
\bibitem{melby04} P. Melby, F. Vega Reyes, A. Prevost, R. Robertson, P. Kumar, D.A. Egolf, and J.S. Urbach, J. Phys. Condens. Mattter, \textbf{17}, S2689 (2005).
\bibitem{jaeger} H. M. Jaeger, S. R. Nagel and R. P. Behringer, Rev. Mod. Phys. \textbf{68}, 1259 (1996).
\bibitem{kadanoff} L. P. Kadanoff, Rev. Mod. Phys. \textbf{71}, 435 (1999).
\bibitem{prevost04} A. Prevost, P. Melby, D.A. Egolf, J.S. Urbach, Phys Rev. E \textbf{70}, 050301(R) (2004).
\bibitem{olafsen05} J. S. Olafsen and J. S. Urbach, {\it Phys. Rev. Lett.}, in press (2005).
\bibitem{kudrolli} A. Kudrolli, Rep. Prog. Phys. \textbf{67}, 209 (2004).
\bibitem{ottino} J. Ottino and D. Khakhar, Annu. Rev. Fluid Mech. \textbf{32}, 55 (2000).
\bibitem{asakura} S. Asakura, F. Oosawa, J. Chem. Phys. \textbf{22} 1255 (1954).
\bibitem{russel} W.B. Russel, D.A. Saville, and W.R. Showalter, \emph{Colloidal Dispersions} (Cambridge University Press, Cambridge 1989).
\bibitem{frydel05} D. Frydel and S. A. Rice, Phys. Rev. E \textbf{71}, 041402 (2005).
\bibitem{mullin} T. Mullin, Phys. Rev. Lett. \textbf{84} 4741 (2000).
\bibitem{reismullin} P.M. Reis, T. Mullin, Phys. Rev. Lett. \textbf{89} 244301 (2002).
\bibitem{reis04} P.M. Reis, G. Ehrhardt, A. Stephenson, and T. Mullin, Europhys. Lett. \textbf{66} 357 (2004).
\bibitem{aumaitre} S. Auma\^itre, C. A. Kruelle, and I. Rehberg, Phys. Rev. E \textbf{64} 041305 (2001).
\bibitem{ehrhardt04} G.C.M.A. Ehrhardt, A. Stephenson, and P.M. Reis, cond-mat 0403273 (2004).
\bibitem{prevost} A. Prevost, D. A. Egolf, and J. S. Urbach, Phys. Rev. Lett. \textbf{89}, 084301 (2002).
\bibitem{nie} X. Nie, E. Ben-Naim, and S. Y. Chen, Europhys. Lett. \textbf{51}, 679 (2000).
\bibitem{Olafsen99} J.S. Olafsen and J.S. Urbach, Phys. Rev. E \textbf{60} R2468 (1999).
\bibitem{rapaport} D. C. Rapaport, \emph{The Art of Molecular Dynamics Simulation} (Cambridge Univ. Press, Cambridge 1995), p. 20.
\bibitem{puglisi} A. Puglisi, A. Baldassarri, and V. Loreto, Phys. Rev. E., \textbf{66}, 061305 (2002).
\bibitem{barrat} A. Barrat, V. Loreto, A. Puglisi, Physica A \textbf{334}, 513 (2004).
\bibitem{deplete-ref} R. Dickman, P. Attard, and S. Simonian, J. Chem. Phys. \textbf{107}, 205 (1997).
\end{thebibliography}
\end{document}